\documentclass[aps,prb,twocolumn,superscriptaddress,floatfix]{revtex4}
\usepackage{color}
\usepackage{graphicx}
\usepackage{epsfig}
\bibstyle{apsrev.bib}

\begin{document}
\input epsf.sty

\title{Critical and tricritical singularities
of the three-dimensional random-bond Potts model for large $q$}

\author{M. T. Mercaldo}
\affiliation{Dipartimento di Fisica ``E.R. Caianiello''  and  Istituto
Nazionale per la Fisica della Materia, Universit\`a degli Studi di
Salerno, Baronissi, Salerno I-84081, Italy}
\author{J-Ch. Angl\`es d'Auriac}
\affiliation{
Centre de Recherches sur les Tr\'es Basses
Temp\'eratures\thanks{U.P.R. 5001 du CNRS, Laboratoire conventionn\'e
avec l'Universit\'e Joseph Fourier}, B. P. 166, F-38042 Grenoble,
France}
\author{F. Igl\'oi}
\affiliation{
Research Institute for Solid State Physics and Optics,
H-1525 Budapest, P.O.Box 49, Hungary}
\affiliation{
Institute of Theoretical Physics,
Szeged University, H-6720 Szeged, Hungary}

\date{\today}

\begin{abstract}
  We study the effect of varying strength, $\delta$, of bond
  randomness on the phase transition of the three-dimensional Potts
  model for large $q$. The cooperative behavior of the system is
  determined by large correlated domains in which the spins points
  into the same direction. These domains have a finite extent in the
  disordered phase. In the ordered phase there is a percolating cluster of correlated spins.
  For a sufficiently large disorder $\delta>\delta_t$ this percolating
  cluster coexists with a percolating cluster of non-correlated spins.
  Such a co-existence is only possible in more than two dimensions.
  We argue and check numerically that $\delta_t$
  is the tricritical disorder, which separates the first- and
  second-order transition regimes. The tricritical exponents are
  estimated as $\beta_t/\nu_t=0.10(2)$ and $\nu_t=0.67(4)$. We claim
  these exponents are $q$ independent, for sufficiently large $q$.
  In the second-order transition regime
  the critical exponents $\beta_t/\nu_t=0.60(2)$ and $\nu_t=0.73(1)$
  are independent of the strength of disorder.
\end{abstract}

\maketitle

\newcommand{\bc}{\begin{center}}
\newcommand{\ec}{\end{center}}
\newcommand{\be}{\begin{equation}}
\newcommand{\ee}{\end{equation}}
\newcommand{\beqn}{\begin{eqnarray}}
\newcommand{\eeqn}{\end{eqnarray}}

\section{Introduction}

The effect of bond randomness on the critical behavior of
ferromagnetic models is well understood if the non-random system has a
second-order phase transition\cite{harris}.  Much less is known,
however, if this transition is discontinuous\cite{imrywortis}. In two
dimensions (2d) rigorous results asserts that for any type of
continuous disorder the transition softens into a second order
one\cite{aizenmanwehr}. Recent numerical studies of the 2d $q$-state
random bond Potts model\cite{Cardy99} (RBPM) have shown that the
magnetization exponent of this model is
$q$-dependent\cite{pottsmc,pottstm} and
saturates\cite{olson99,jacobsenpicco} for large-$q$ at a possibly
exactly known value\cite{ai03,long2d}. On the other hand the energy
exponent, $\nu$, is found to show only a very weak variation with $q$.

In real 3d systems the effect of bond randomness is more complex and
here we are lacking rigorous results. It is demonstrated
experimentally that the isotropic to nematic transition of $nCB$
liquid crystal turns to second order for sufficiently strong
disorder\cite{exp}.  The same type of softening effect is found in
Monte Carlo simulations for the $q=3$ and $q=4$ Potts models, both for
site\cite{uzelac,pottssite} and bond dilution\cite{pottsbond}.  With
very large computational effort it was possible to locate the second
order transition point and to estimate the critical exponents, which
are found $q$ dependent both for the magnetization and for the energy
density.

The first-order transition regime (for weak disorder) and the
second-order transition regime (for strong disorder) are
separated by a tricritical point the properties of which are
conjectured to be related to the critical point of the random
field Ising model (RFIM). As shown by Cardy and Jacobsen\cite{pottstm} in the limit
of $d \to 2$ and $q \to \infty$ the interface Hamiltonian of the two
problems have the same type of solid-on-solid (SOS) description, from
which follows that the energy exponents of the tricritical RBPM are
equivalent to the magnetization exponents of the critical RFIM.
Furthermore, analyzing the renormalization group (RG) flow it was conjectured
that the above mapping stays valid for $d > 2$, in particular at $d=3$. 
Thus the tricritical exponents should be $q$ independent, at least for
large $q$, leading to the same exponents for
any disorder induced tricritical points.

These conjectures, which could be of experimental relevance, have not
yet been verified numerically. The inaccuracies in the simulations
have mainly two sources: i) it is difficult to precisely locate the
tricritical point due to strong cross-over effects and ii) for not too
large $q$ the tricritical disorder is quite small which results in
large breaking-up lengths in the system\cite{long2d}, 
thus one has to treat quite large lattices.

In the present paper we consider the 3d ferromagnetic Potts model\cite{Wu} for
large value of $q$, which has a strongly first-order transition for
non-random couplings and study the effect of bond randomness including
also the case of bond dilution. In the large-$q$ limit the
high-temperature expansion of this problem\cite{kasteleyn} is dominated by a single
diagram\cite{JRI01} which is exactly calculated by a combinatorial optimization
method\cite{aips02}. Weak disorder is studied perturbatively, whereas for stronger
disorder we have performed extensive numerical calculations. In
particular we have studied the properties of the tricritical point and
checked its possible relation with the critical fixed point of the
RFIM. We have also studied the form and universality, i.e. disorder
independence, of the critical singularities. For theses quantities
some results have already been announced in a letter\cite{mai05}.

The structure of the paper is the following. The RBPM and the
optimization method used in the study for large $q$ is presented in
Sec.\ref{sec_model}. Perturbative treatment of the problem in the weak
disorder limit is shown in Sec.\ref{weak}. Numerical results about
the phase diagram as well as on critical and tricritical singularities
are given in Sec.\ref{sec_numerics} and discussed in
Sec.\ref{sec_disc}.

\section{Model and its phase diagram}
\label{sec_model}

The $q$-state Potts model\cite{Wu} is defined by the Hamiltonian:
\begin{equation}
\mathcal{H}=-\sum_{\left\langle i,j\right\rangle }J_{ij}\delta(\sigma_{i},\sigma_{j})
\label{eq:hamilton}
\end{equation}
in terms of the Potts-spin variables, $\sigma_{i}=0,1,\cdots,q-1$. Here
$i$ and $j$ are sites of a cubic lattice and the summation runs over nearest neighbors.
The couplings, $J_{ij}>0$, are ferromagnetic and identically and independently distributed
random variables. In this paper we use a bimodal distribution:
\begin{equation}
P(J_{ij})= p\delta(J(1+\delta)-J_{ij})+(1-p)\delta(J(1-\delta)-J_{ij}) \label{eq:bimodal}
\end{equation}
with $p=1/2$. The parameter $0<\delta \le 1$ plays the role of the strength of disorder and
at $\delta=0$ and $\delta=1$ we recover the non-random and
the diluted systems, respectively.

For a given set of couplings the partition function of the system is conveniently to write in
the random cluster
representation\cite{kasteleyn} as:
\begin{equation}
Z =\sum_{G}q^{c(G)}\prod_{ij\in G}\left[q^{\beta J_{ij}}-1\right]
\label{eq:kasfor}
\end{equation}
where the sum runs over all subset of bonds, $G$ and $c(G)$ stands for
the number of connected components of $G$. In Eq.(\ref{eq:kasfor}) we use
the reduced temperature, $T \to T \ln q$ and its inverse $\beta
\to \beta/\ln q$, which are of $O(1)$ even in the large-$q$ limit\cite{long2d}. In this limit we have
$q^{\beta J_{ij}} \gg 1$ and the partition function can be written as
\begin{equation}
Z=\sum_{G\subseteq E}q^{\phi(G)},\quad \phi(G)=c(G) + \beta\sum_{ij\in G} J_{ij}\label{eq:kasfor1}
\end{equation}
which is dominated by the largest term, $\phi^*=\max_G \phi(G)$. 
Note that the optimal set itself
generally depends on the temperature. The
free-energy per site is proportional to $\phi^*$ and given by $-\beta
f= \phi^*/N$ where $N$ stands for the number of sites of the lattice.

\begin{figure}
  \begin{center}
     \includegraphics[width=2.35in,angle=0]{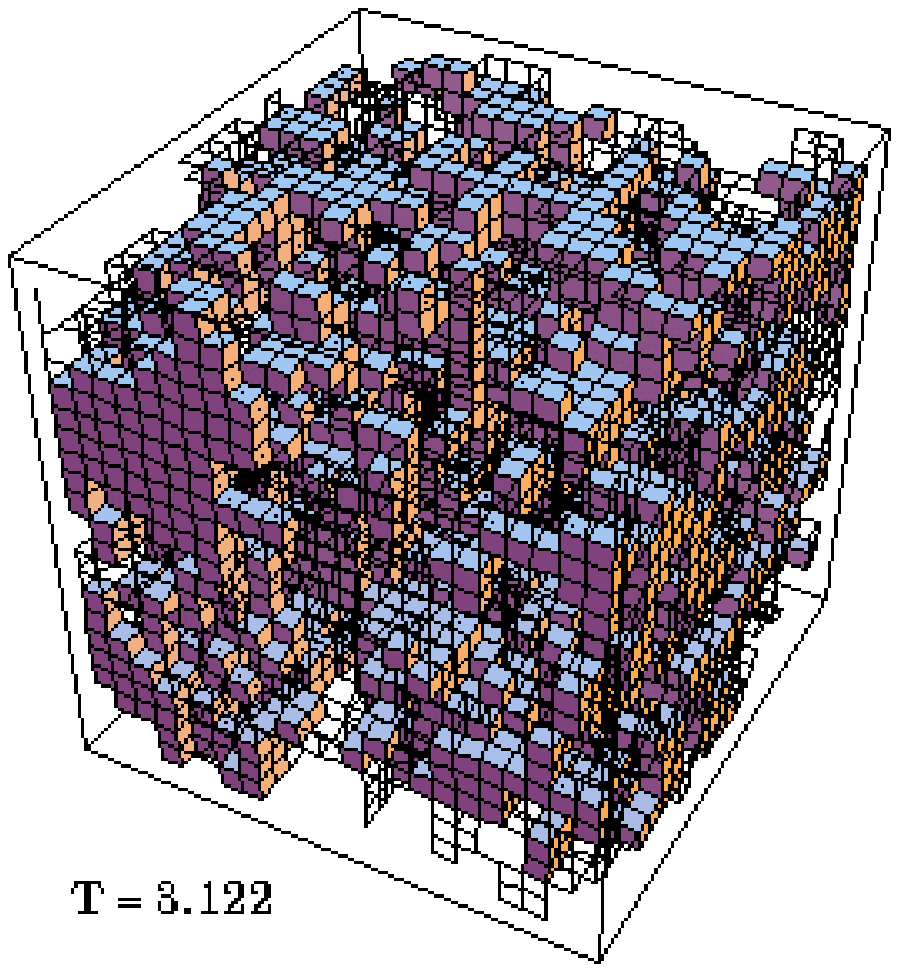}
     \includegraphics[width=2.35in,angle=0]{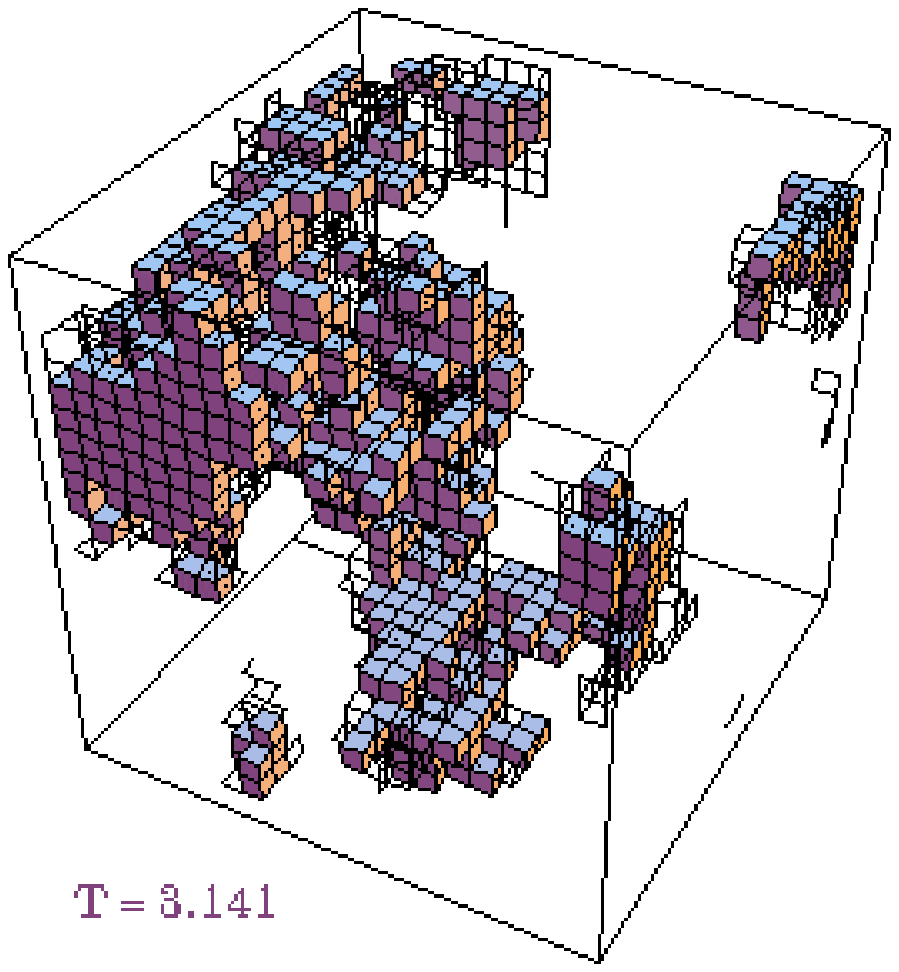}
     \includegraphics[width=2.35in,angle=0]{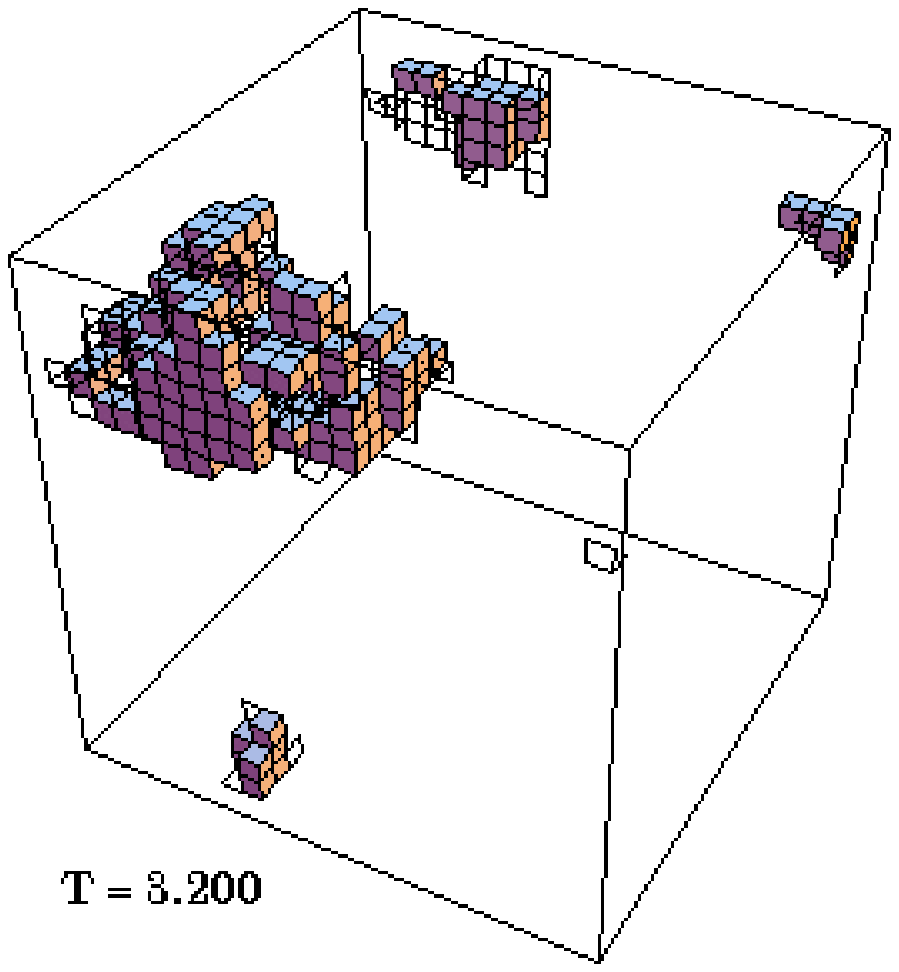}
   \end{center}
   \caption{Connected parts of typical optimal sets with $L=24$ at $\delta=0.875$.Top: $T=3.122J <T_c$,
centre:  $T=3.141 J \approx T_c$ and  bottom: $T=3.200J >T_c$.
}
   \label{Abb1}
 \end{figure}

The optimization problem in Eq.(\ref{eq:kasfor1}) contains a
cost-function, $\phi(G)$, which is sub-modular\cite{gls81} and there is an
efficient combinatorial optimization algorithm, which at any
temperature, works in strongly polynomial time\cite{aips02}. This algorithm
finds a set of bonds which minimizes the cost-function. We call
such a set an optimal set.
The variation of the optimal set with the
temperature is illustrated\cite{movie} in Fig.\ref{Abb1} for $\delta=0.875$ and
$L=24$. At low temperature the optimal graph is compact and the
largest connected subgraph contains a finite fraction of the sites. In
the other limit, for high temperature, most of the sites in the
optimal set are isolated and the connected clusters have a finite
extent, the typical size of which is used to define the correlation
length, $\xi$.

\begin{figure}
  \begin{center}
     \includegraphics[width=2.35in,angle=270]{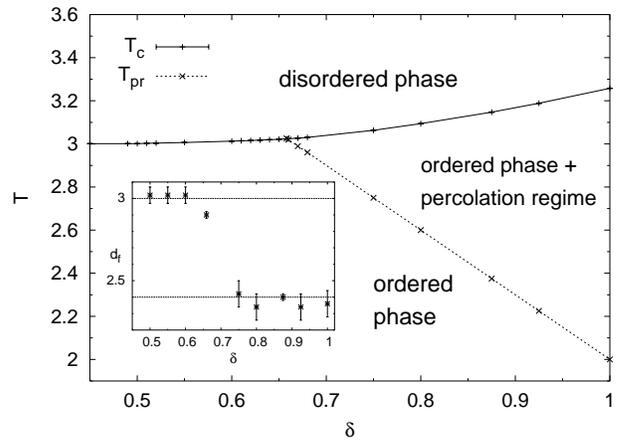}
   \end{center}
   \caption{Phase diagram of the random-bond Potts model for bimodal disorder.
In the ordered phase in the optimal
     set the non-connected sites are percolating for $\delta>\delta_{pr}$
     and $T_{pr}(\delta)<T<T_c(\delta)$. Numerical results indicate, that
     the first- and second-order transition regimes are separated by a
     tricritical disorder, $\delta_{c}$, which corresponds to the
     border of the percolation regime, $\delta_{c}=\delta_{pr}$. Inset:
fractal dimension of the giant connected cluster at the transition
point as calculated in Sections \ref{sec_critic} and \ref{sec_tricritic}. The straight
lines at $d_f=3$ and $d_f=2.40$ indicating the values at the first-order and the second-order
transition regimes,respectively, are guide to the eye. }
   \label{Abb2}
 \end{figure}

Between the low-temperature (ordered) and the high-temperature
(disordered) phases in the thermodynamic limit there is a sharp phase
transition. The numerically calculated phase-diagram as a function of
the temperature, $T$, and the disorder, $\delta$, is shown in Fig.
\ref{Abb2}. A detailed analysis of the phase-diagram is postponed to
Sec.\ref{sec_numerics}. Here we just note that the
transition is of first order for weak disorder, in which case $\xi$
stays finite at the transition point, but the transition is of second
order for strong enough disorder, when the correlation length is
divergent at the transition point. This second possibility is
illustrated in the central part of Fig.\ref{Abb1}, when at the transition point the
largest connected cluster is a fractal and its fractal dimension,
$d_f$, is related to the magnetization critical exponent, see in
Sec.\ref{sec_numerics}. The fractal dimension of the giant connected
cluster at the transition point is shown in the inset of Fig. \ref{Abb2}, as calculated in
Sec.\ref{sec_critic}. In the first-order regime it is $d_f=3$, i.e.
the cluster is compact, whereas in the second-order regime it is
$d_f<3$ and practically independent of the strength of disorder.  At
the tricritical disorder, $\delta_t$, the giant cluster is a fractal
but its fractal dimension is different from that in the critical
regime, see in Sec.\ref{sec_tricritic}.

\section{Perturbative calculation for weak disorder}
\label{weak}

\subsection{Non-random model and Imry-Ma type argument for weak disorder}

In the non-random model, $\delta=0$, there are only two homogeneous
optimal sets, which correspond to the $T=0$ and $T \to \infty$
solutions, respectively, see in Ref.\cite{movie}. For $T<T_c(0)$ it is the
fully connected diagram with a free-energy: $-\beta N f=1+N\beta J d$
and for $T>T_c(0)$ it is the empty diagram with $-\beta N f=N$.
Consequently the transition point is located at: $T_c(0)=Jd/(1-1/N)$
and the latent heat is $\Delta e/T_c(0)=1-1/N$.

In the presence of disorder, $\delta>0$, the optimal sets are
homogeneous only in a limited temperature range and new
non-homogeneous optimal diagrams appear, see in Fig.\ref{Abb1}. In the
disordered phase in the vicinity of $T_c(0)$ the typical linear size of the
connected clusters, $l$, can be estimated along the lines of the
Imry-Ma argument\cite{imryma}. Adding to the optimal set a cluster with a number of sites
of the order $l^d$ will decrease the number of
connected components, and therefore increase the cost-function by $l^d$.
On the other hand, it will add a number of the order $dl^d-dl^{d-1}$
of bonds, each having a weight $(1 \pm \delta)/d$ since the 
temperature is close to $T_c(0)$. It results in a competition
between a term behaving like $l^{d/2}$, which represents the gain due
to disorder fluctuations and
a term like  $l^{d-1}$, which is the lost due to the creation
of an interface.
In $d=2$ the two terms
balance each other and extreme fluctuations of the disorder will
create clusters of unlimited size\cite{long2d}. 
As a consequence in 2$d$ there is
no coexistence of pure phases and the transition is of second order
for any small amount of disorder\cite{aizenmanwehr}.

On the contrary in $d>2$, in particular in three dimensions for weak
disorder the surface term is dominating, thus the connected clusters
have a finite extent and the transition is of first-order in
accordance with the phase-diagram in Fig. \ref{Abb2}. In this case
connected clusters are created due to extreme fluctuations of disorder, and the
statistics of these rare regions will be
considered in the following subsection. A similar analysis of the
ordered phase will be given afterward in Sec. \ref{isolated}.

\subsection{Disordered phase}
\label{connected}
Here we make the estimate of the free-energy as simple as possible and
therefore we consider rare regions of the shape of cubes of linear
linear size, $l \ge 2$, in which all the $n_+=3(l^3-l^2)$ internal
couplings are strong, being $J(1+\delta)$. The cost-function of the
diagram in which all the bonds of the cube are present relative to the
empty graph is given by:
\begin{equation}
\Delta f_+(l)=3\beta(l^3-l^2)J(1+\delta)-(l^3-1)\;.
\label{Delta_f}
\end{equation}
The optimal set is then the inhomogeneous one containing the connected
cluster, if $\Delta f_+(l)>0$, which could take place if the disorder
satisfies the relation, $\delta > \delta^+(l)$, with
\begin{equation}
\delta_+(l)=\frac{l+1}{l^2}\;.
\end{equation}
In another words for a given disorder, $\delta$, there is a limiting
size, $l_+(\delta)=(1+\sqrt{1+4 \delta})/2 \delta \approx 1/\delta$,
and only large enough clusters with $l \ge l_+(\delta)$ can exist in
the optimal graph. provided the temperature is in the range of
$3<T/J<T_+(l)/J=3+3(\delta-\delta_+(l))/(1+\delta_+(l))$.  Thus for
weak disorder only large (and very rare) clusters can be found and as
the disorder is increased at discrete values of $\delta$ new, smaller
and more probable connected clusters will appear.  This mechanism will
lead to discontinuous behavior of the free energy as a function of the
disorder, which is indeed observed in numerical calculations.

The probability of having all the $n_+$ couplings strong in a cube is
exponentially small, $P_+(l)=2^{-n_+}$, thus the density of connected
cubes of size $l$ in the optimal set is given by
$\rho_+(l)=2^{-3(l^3-l^2)}$. The free energy of the non-homogeneous
optimal set is given by the sum of contributions of the connected
clusters of different size:
\be
-\beta N f_+ \simeq N + N \sum_{l \ge l_+(\delta)} \Delta f_+(l) \rho_+(l)\;.
\label{f+}
\ee
Since $\rho_+(l)$ is a very rapidly decreasing function of $l$ the sum
in Eq.(\ref{f+}) is dominated by the term with $l=l_+(\delta)$.

\subsection{Ordered phase}
\label{isolated}
In the ordered phase, $T \le T_c(0)$, we consider rare regions of
extreme disorder fluctuations also of the shapes of cubes with $l \ge
1$, so that all the couplings, $n_-(l)=3(l^3+l^2)$, starting from the
points of the cubes are weak, being $J(1-\delta)$.  The cost-function
of the diagram in which all weak bonds of the cube are absent, i.e.
there is a cube of isolated points embedded into the full diagram,
relative to the full graph is given by:
\begin{equation}
\Delta f_-(l)=l^3-3\beta(l^3+l^2)J(1-\delta)\;.
\label{Delta_f-}
\end{equation}
As for the disordered phase the optimal set is the inhomogeneous one
containing the isolated points, if $\Delta f_-(l)>0$, which is the
case for strong enough disorder, $\delta > \delta_-(l)$, with
\begin{equation}
\delta_-(l)=\frac{1}{l+1}\;.
\end{equation}
Now for a given disorder, $\delta$, the limiting size is,
$l_-(\delta)=1/ \delta -1 \approx 1/\delta$, and the $l \ge
l_-(\delta)$ ``empty clusters'' exist in the optimal graph in the
range of temperature
$3>T/J>T_-(l)/J=3+3(\delta-\delta_-(l))/(1+\delta_-(l))$.  Now the
density of cubes of isolated points of size $l$ in the optimal set is
given by $\rho_-(l)=2^{-3(l^3+l^2)}$ and the free energy of the
non-homogeneous optimal set can be written as:
\be
-\beta N f_- \simeq1 + N\beta J d + N \sum_{l \ge l_-(\delta)} \Delta f_-(l) \rho_-(l)\;,
\label{f-}
\ee
which is dominated by the term with $l=l_-(\delta)$.

\subsection{Phase transition}
\label{phase-t}
The phase-transition temperature, $T_c(\delta)$, is obtained by
equating the free-energy in the two phases: $f_-=f_+$. Since it is
comparatively easier to create an ``empty cluster'' in the ordered
phase, than a connected cluster in the disordered phase which results
in the shift of the phase boundary towards higher temperatures as the
disorder is increasing, see Fig. \ref{Abb2}. Further observation, that
the phase-boundary is discontinuous at $\delta_+(l)$ and
$\delta_-(l)$, for all integer $l$-s. For small $\delta$, which
corresponds to large $l$-s these jumps are very frequent and in a
mathematical point of view the phase boundary is a non-analytical
function of $\delta$. However these jumps are very small and the
phase-boundary can be approximated by a continuous curve which
asymptotically behaves as:
\begin{equation}
\ln (T_c-T_c(0)) \sim -\frac{1}{\delta^3}\;.
\label{T_shift}
\end{equation}
We obtain similarly for the latent heat and for the jump of the
magnetization at the transition point:
\begin{equation}
\ln (\Delta e/T_c-1) \sim \ln( \Delta m-1) \sim-\frac{1}{\delta^3}\;.
\end{equation}
Thus the phase transition stays first order and as the disorder is
switched on there is an essential singularity in the thermodynamical
quantities as a function of $\delta^{-3}$.

\subsection{Distribution of the finite-size transition temperature}
\label{sec_distr}
For a large finite system of linear size, $L$, and for a given
realization of the disorder one can define a transition temperature,
$T_c(L)$, at which the disordered phase and the ordered phase (having
a spanning giant cluster) coexists. The distribution of $T_c(L)$ is discontinuous
for a finite system but expected to be
Gaussian for large $L$, so that characterized by its average,
$T^{av}_c(L)$, and its variance, $var[T_c(L)]=[\Delta T_c(L)]^2$. The
shift of the average is asymptotically given by:
\be
T^{av}_c(L)-T_c(\infty) \sim L^{-1/\tilde{\nu}}\;,
\label{T_c_L}
\ee
whereas the width scales with another exponent, $\nu$, as:
\be
\Delta T_c(L) \sim L^{-1/{\nu}}\;.
\label{DT_c_L}
\ee
These relations define the exponents $\nu$ and $\tilde{\nu}$ and hold
for second order\cite{wd95,ah96,psz97,wd98,ahw98} as well as for first
order phase transitions\cite{fisher,mg05}. At a first-order transition
what we study here $\tilde{\nu}$ is just the discontinuity fixed point
value of the correlation length exponent\cite{nn,fb} and given by
$\tilde{\nu}=\nu_d=1/d$. This exponent describes the variation of a
diverging length-scale which can be measured in a {\it typical
  sample}. The width of the distribution can be obtained from the
consideration that the density of clusters, $\rho_{\pm}$, has a
fluctuation in finite systems which has a width of, $\Delta
\rho_{\pm}(L)/\rho_{\pm} \sim \rho L^{-d/2}$, according to the central
limit theorem. The fluctuations in $\Delta \rho_{\pm}(L)$ are then
seen in the fluctuations of $T_c(L)$, too, leading to an exponent,
$\nu=2/d$. This exponent is related to another diverging length-scale,
which can be deduced from the study of {\it average} quantities. Note
that this length is not present in the pure system and by switching on
disorder its prefactor, $\rho_{\pm}$, presents an essential
singularity as a function of $\delta$. These results are in accordance
with the considerations presented in Sec.VII of
Ref.[\onlinecite{fisher}]. Numerical analysis of $T_c(L)$ in for
stronger disorder is given in Sec.\ref{sec_numerics}.

\subsection{Breaking-up lengths}
\label{sec_breaking}
For weak disorder the density of elementary excitations is very small
and the average distance between two excitations is given by $L_+^b
\sim \rho_+(\delta)^{-1/3}$ and $L_-^b \sim \rho_-(\delta)^{-1/3}$, in
the two phases, respectively. $L_+^b$ and $L_-^b$ can be interpreted
as breaking-up lengths, since in a finite system of linear size,
$L<L_{\pm}^b$, the optimal set is homogeneous. Using results about the
critical densities we obtain:
\beqn
\log_2 L_+^b=(1+2\delta-2\delta^2+\sqrt{1+4\delta})/2\delta^3,
\nonumber\\
\log_2 L_-^b=\left(\delta^{-1}-1\right)^3+\left(\delta^{-1}-1\right)^2\;,
\label{break}
\eeqn
which in principle is valid only at $\delta=\delta_{\pm}(l)$, however
as an interpolation formula we can use them for not too small $\delta$-s, too.

\begin{table}
\caption{Breaking-up disorder in the ordered phase: comparison of the interpolation
formula in Eq.(\ref{break}) with numerical results \label{table:1}}
 \begin{tabular}{|c|c|c|}  \hline
  $L$  & $\delta^b_{int}$ & $\delta^b_{num}$ \\ \hline
  $7$  & $ 0.466 $ & $0.469$ \\
  $9$  & $0.454$ & $0.458$ \\
  $15$ & $0.434$ & $0.438$  \\ 
  $17$  & $0.430$& $ 0.435$ \\
  $23$ & $0.420$ &$ 0.429$ \\
  $31$  & $ 0.412$&$  0.423$ \\
  $39$ & $0.406$ & $0.418$ \\

 \hline
  \end{tabular}
  \end{table}
  
  It is easy to see that the braking-up lengths in the ordered phase
  are much smaller, then in the disordered one. For example at the
  value of $\delta=1/2$, which corresponds to the existence of a
  single hole in the ordered phase and to a $l=2$ connected cube in
  the disordered phase, respectively, $L_-^b=4$, whereas $L_+^b
  \approx 2^{13}$.  We have checked the accuracy of the formula in
  Eq.(\ref{break}) by comparing the predicted breaking-up disorder,
  $\delta_-^b$, for a given size, $L$, with that calculated
  numerically. As seen in Table\ref{table:1} there is a satisfactory
  agreement, even for not too small $\delta$-s.

\section{Numerical results}
\label{sec_numerics}

In the numerical calculation we have treated samples with random
couplings having a cubic shape with periodic boundary conditions and a
linear size $L=16,~24,~32$. For $\delta=0.875$ we went up to $L=40$ and in
some cases we made calculations for odd values of $L$, too. The
free-energy
as well as the magnetization is calculated exactly by the
combinatorial optimization algorithm, called as ``optimal
cooperation''\cite{aips02} and averaging is performed over several thousands of
samples, for the largest size the number of realizations was several
hundreds.  For a fixed temperature the optimal cooperation algorithm
works in strongly polynomial time. As we have already mentioned in the
application of the method in $2d$\cite{long2d} for a finite system the free-energy
is a piece-wise linear function of the temperature.
For the bimodal disorder the number of linear parts, $N_p$, is found to 
increase with the size of the system $L$
as well as with the disorder strength $\delta$.
A rough estimation yields
\be
N_p \simeq C L^{1+\delta}\;,
 \label{hyp}
\ee
with the constant $C$ around one half.
We have managed to implement our method in such
a way that we can calculate the free-energy in the whole temperature
range, i.e. in all linear parts. 
Note that with the hypothesis (\ref{hyp})
the exact calculation
of the free energy {\em in the whole temperature range} is still
polynomial, whatever the value of the disorder $\delta$, 

As we have shown in Sec.\ref{sec_breaking} the breaking-up length is
large for weak disorder and it is of the order of $L^b=40$ for $\delta
\approx 0.4$. Therefore we have restricted ourselves to the disorder
range: $0.4 < \delta \le 1$, which contains all interesting parts of
the phase diagram.

\subsection{Phase diagram}
\label{sec_phased}

\begin{figure}
  \begin{center}
     \includegraphics[width=2.35in,angle=270]{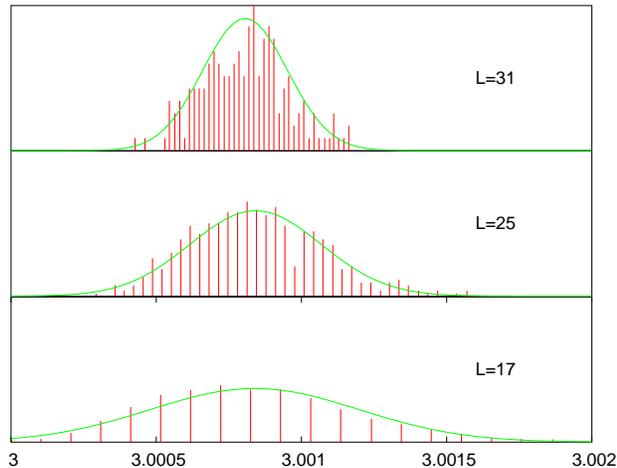}
   \end{center}
   \caption{Distribution of the finite-size percolation transition temperatures in the first-order
transition regime at $\delta=.5$ for three different sizes.
The Gaussian distributions having the same average and
variance are normalized and serve as a guide to the eye..
}
   \label{Abb3}
 \end{figure}

\begin{figure}
  \begin{center}
     \includegraphics[width=2.35in,angle=270]{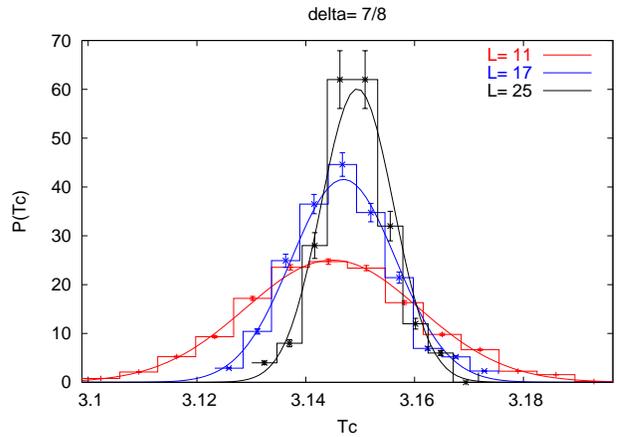}
   \end{center}
   \caption{The same as in Fig.\ref{Abb3} in the second-order transition regime at $\delta=0.875$}.
   \label{Abb4}
 \end{figure}

The numerically calculated phase diagram as a function of disorder and
temperature is already presented in Sec.\ref{sec_model} in Fig.
\ref{Abb2}, here we give a detailed analysis of the results. The phase
boundary between the ordered and disordered phases is calculated by
considering two quantities. Finite-size or percolation transition
temperatures, $T_c(L)$, are identified in each sample at the point
where the largest connected cluster of the optimal set starts to
percolate the finite sample. The distribution of $T_c(L)$ is shown in
Fig.\ref{Abb3} in the first-order transition regime ($\delta=0.5$) and in
Fig.\ref{Abb4} in the second-order transition regime ($\delta=0.875$).

 While for $\delta=0.875$ the distribution of the transition
 temperatures is well described by a Gaussian even for relatively
 small systems, for $\delta=0.5$ the distributions show deviations
 from a Gaussian.  As we can see in Fig.\ref{Abb3} in the
 first-order regime the distribution for finite $L$ consists of well
 separated peaks and it becomes (quasi-)continuous only in the
 thermodynamic limit. Also the distributions are non-symmetric and the
 skewness seems to vanish slowly with the size of the system.  It is
 also evident from Figs.\ref{Abb3} and \ref{Abb4} that the finite-size shift of the
 average transition temperature has a different sign in the two regimes and it is much smaller
for a first-order transition in particular if we compare with the variation of the width
of the distributions. On the contrary in the second-order transition regime
these two characteristics of the distribution are in the same order, see in
 Fig.\ref{Abb4}. These observations are in accordance with the
 scaling picture in Sec. \ref{sec_distr}, which predicts
two different exponents, $\tilde{\nu}$ and $\nu$  in the first-order transition
regime, whereas in the second-order regime these exponents are the same. However, due to the
 comparatively small sizes we have and the deviations of the distributions from the Gaussian we
 could not numerically estimate these exponents.

We have also calculated the transition temperature from the position
of the maximum of the averaged specific heat, $C_v(T,L)$,
which is expected to be shifted in finite systems as the maxima of
$T_c(L)$, as given in Eq.(\ref{T_shift}). Indeed the numerical results
on the specific heat in Fig. \ref{Abb9}  at $\delta=0.875$ are compatible with
the same estimate for $T_c$, as obtained in Fig.\ref{Abb4} through the
percolation transition temperatures.

As discussed before magnetic order in the system is related to the properties
of the largest connected cluster in the optimal set: it has a finite
extent in the disordered phase, whereas a finite fraction of sites
belongs to this cluster in the ordered phase. The optimal set in the
ordered phase, however, can be of two different kinds as far as the
structure of isolated points is considered. For weak disorder,
$\delta < \delta_{pr}$, or for low temperature, $T < T_{pr}$, the
isolated sites form finite clusters. This is always the case in $2d$.
In $3d$, however, and this is a new feature of these systems for
strong enough disorder and high enough temperatures the isolated sites
percolate the sample, too. This percolation regime is also indicated
in Fig.\ref{Abb2} and we argue below that its existence close to $T_c$
is necessary to have a second-order transition in the system. Indeed,
the correlation length in the ordered phase is given by the size of
the largest connected finite cluster, which is isolated from the giant
connected cluster. Since this large finite cluster is embedded into
isolated points its size can be divergent at $T_c$ only if the
isolated sites percolate the sample. In this way we obtain for the
value of the tricritical disorder, $\delta_t$, separating the first-
and second-order transition regimes that it satisfies the relation,
\be
\delta_t \ge \delta_{pr}\;.
\label{delta_t}
\ee
Numerical results which are presented below are in favor of the
conjecture that in this relation the equality holds.

\begin{figure}
  \begin{center}
     \includegraphics[width=2.35in,angle=270]{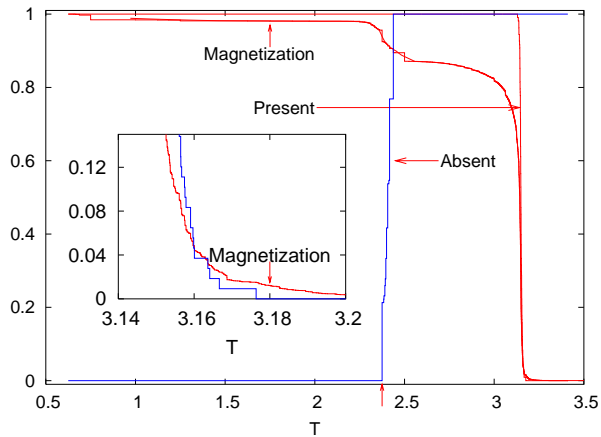}
   \end{center}
   \caption{Magnetization and percolation probabilities of the optimal
     set (``Present'': for the occupied bonds of the giant cluster;
     ``Absent'': for the isolated sites) of the random model with
     $\delta=0.875$, as a function of the temperature in a finite
     lattice with $L=24$. The arrow at $T = T_{pr}=2.375J$ indicates
     the percolation temperature in Eq.(\ref{T_pr}) and for $T <
     T_{pr}$ the magnetization is close to that for ordinary
     percolation. The singular jumps in the magnetization are a
     consequence of the discrete form of the probability distribution.
     Inset: in the vicinity of the transition point the magnetization
     and the percolation probability of the present bonds are close to
     each other, the differences are due to finite-size effects.}
   \label{Abb5}
 \end{figure}

For the percolating transition temperature of the isolated sites,
$T_{pr}(\delta)$, we make the following calculation. In the dilute model,
$\delta=1$, all the strong bonds, $J_1=2J$, are present in the optimal
set, provided the temperature is below the value of $J_1$. For $T >
J_1$, however, in the optimal set there are no dangling bonds, i.e. by
increasing the temperature over $T=J_1$ sites which have just one
strong bond are removed from the optimal set. Since the dangling bonds
are a finite fraction of the bonds\cite{staufferaharony} the non-connected sites become
percolating at $T_{pr}(1)=2J$. For $\delta < 1$ there are also weak
$J_2$ bonds in the system and the removal of one strong dangling bond
from the optimal set is accompanied by the removal of some weak bonds
at the same time. The average number of removed decorating weak
bonds is four which is possible in the
temperature range:
\be
T_c>T>T_{pr}(\delta)=J_1+4J_2=J(5-3\delta)\;.
\label{T_pr}
\ee
The numerical results show that at $T_{pr}(\delta)$ in a finite
fraction of samples there is a giant cluster of isolated points which
spans the finite cube, see in Fig.\ref{Abb5}.

 At the percolation transition temperature there is a sudden change in
 the structure of the giant connected cluster, which results in
 singularities in the thermodynamical quantities. As an illustration
 we show in Fig. \ref{Abb5} the temperature dependence of the
 magnetization at $\delta=0.875$, i.e. in the second-order transition
 regime.  The magnetization goes to zero at the phase-transition
 point, which - in the thermodynamic limit - coincides with the
 percolation transition of the occupied bonds. For finite systems the
 magnetization and the probability of having a spanning cluster has
 small deviations, as illustrated in the inset of Fig. \ref{Abb5}.
 The magnetization has another singular jumps for $T<T_c$, which are
 due to the discrete form of the disorder. Among these singularities
 the most pronounced is that around the percolation temperature at
 $T_{pr}=19/8$. As seen in Fig. \ref{Abb5} the percolation probability
 of the isolated sites has a finite value at $T_{pr}$ and it goes to the
value of  one within a small temperature range. It is expected that by
 decreasing the disorder to the tricritical value, the singularities
 at $T_c$ and $T_{pr}$ merge into a new type of tricritical
 singularity.

\subsection{Order of the transition}

To decide about the order of the transition one generally studies the
behavior of the latent heat in the system. This type of analysis,
however, for the large-$q$ state Potts model is complicated, if the
disorder is discrete, as in our case. As already discussed in 2d\cite{long2d} the
internal energy of the system displays discontinuities, both at and
outside the critical point.  These discontinuities, however, are
generally connected to such degeneracies of the optimal set which are
related to the removal of finite number of bonds, thus do not modify
the global structure of the giant connected cluster. As a consequence
these singularities can not be interpreted as a sign of first-order
transition. In $3d$ and for the bimodal disorder used in this paper
this type of non-generic discontinuities of the internal energy are
also present therefore we did not try to make an analysis of this
quantity.

\begin{figure}
\includegraphics[width=2.35in,angle=270]{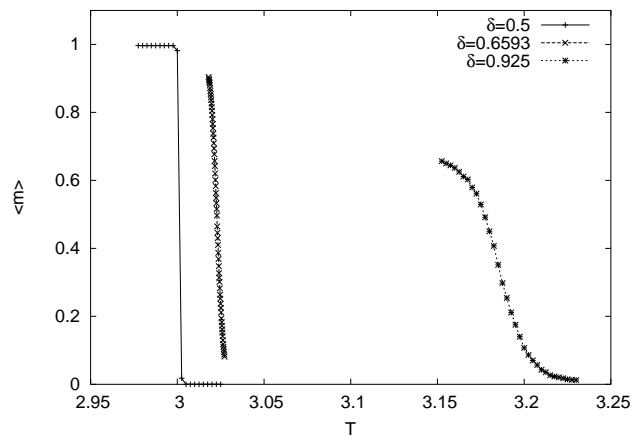}
 \caption{Temperature dependence of the magnetization in the transition region for $L=16$ with
   different strength of disorder. From left to right
   $\delta=0.5$ - first-order regime; $\delta_t=0.6593$ - tricritical point;
$\delta=0.925$ - second-order regime.}
 \label{Abb6}
\end{figure}

Instead we have studied the magnetization in the system, which is
defined as the fraction of sites in the largest connected cluster. A
jump in the magnetization indicates a fundamental change in the shape
of the largest cluster, thus a first-order transition. The
magnetization as a function of temperature is shown in Fig.\ref{Abb5} for
different values of the disorder corresponding to the different transition
regimes. A direct measurement of the jump in the magnetization is possible
only up to $\delta \leq 0.6$, see in Ref.\cite{mai05}, whereas the lower
limit of the second-order transition regime is estimated through the calculation
of the fractal dimension of the giant cluster at the transition point, see in
Fig.\ref{Abb2}. This type of analysis at $\delta_{pr}$ (see in Sec. \ref{sec_tricritic})
has given a fractal dimension, $d_f^t<3$, which together with Eq.(\ref{delta_t})
indicates that the first-order transition regime stops at $\delta_{pr}$.

\subsection{Second-order transition regime}
\label{sec_critic}

In the second-order transition regime we have made detailed
calculations at the disorder $\delta=0.75,~0.8,~0.875,~0.925$ and at
$\delta=1$. The most detailed studies are performed at
$\delta=0.875$, in which case the largest system is $L=40$ for the
other cases we went up to $L=32$. With these investigations our aim
was to check universality of the critical properties of the system.

\begin{figure}
  \begin{center}
     \includegraphics[width=2.35in,angle=270]{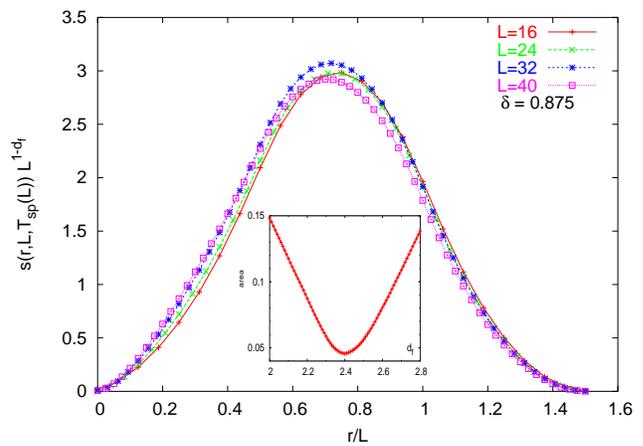}
   \end{center}
   \caption{Scaling plot of the mass of a shell of the infinite cluster, for each size at the average
spanning temperature, see text, at $\delta=0.875$. Optimal collapse is obtained with a
fractal dimension, $d_f=2.40$.
Inset: area of the collapse region as a function of $d_f$.}
   \label{Abb7}
 \end{figure}

The magnetization exponent, $\beta$, and the magnetization scaling
exponent, $x=\beta/\nu$, is related to the fractal dimension of the
giant cluster, $d_f$, as $d-x=d_f$. Among these critical parameters it
is the fractal dimension which can be determined with the highest
precision.  To obtain the fractal dimension we considered a reference
point of the percolating cluster and measured the mass (number of
points) in a shell around the reference point with unit width and
radius, $r$. The average mass, $s(r,L,T)$, is expected to scale close
to the transition point: $t=(T-T_c)/T_c \ll 1$ as:
\begin{equation}
s(r,L,t)=L^{d_f-1} \tilde{s}(r/L,tL^{1/\nu})\;.
\label{s}
\end{equation}
Generally, one sets the second argument of the scaling function,
$\tilde{s}(\rho,\tau)$, to be zero by performing the calculation at
the transition temperature, $T=T_c$. In our case $T_c$ is not known
exactly, therefore we have used another strategy. For each size we
set, $T=\overline{T_c}(L)$, which is the average finite-size
(percolation) temperature at the given size. With this choice the
second argument of the scaling function, $\tau$ is asymptotically
constant and thus the scaling function depends only on one parameter:
$\tilde{s}=\tilde{s}(r/L)$. Our scaling picture is checked in Fig.
\ref{Abb7}, in which the scaling plot of the mass in the shell is
shown for $\delta=0.875$.  The accuracy of the scaling collapse is
measured by the area of the collapse region which is shown in the
inset of Fig. \ref{Abb7}. It is seen that the optimal scaling collapse
is obtained with a fractal dimension, $d_f=2.40(2)$. This type of
analysis of the fractal dimension is repeated for another values of
the disorder, too. Since in these cases the available sizes of the
systems are comparatively smaller we have somewhat larger errors. The
fractal dimensions are shown in the inset of Fig. \ref{Abb2}.

\begin{figure}
  \begin{center}
     \includegraphics[width=2.35in,angle=270]{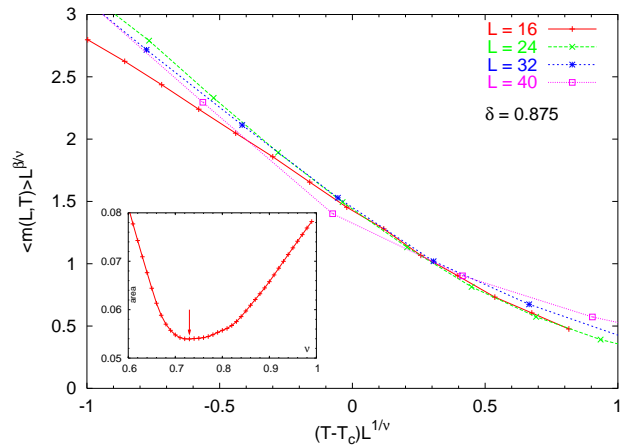}
   \end{center}
   \caption{Scaling plot of the magnetization at $\delta=0.875$ as a
     function of the distance from the critical temperature. By fixing
     $\beta/\nu=d-d_f=2.4$ the best collapse is obtained with
     $\nu=.73(2)$.  In the inset the area of the collapse region is
     shown as a function of $\nu$, the arrow indicates the position of
     the minima.}
   \label{Abb8}
 \end{figure}

 The correlation length critical exponent, $\nu$, can be calculated
 from the shift of the finite-size critical temperature, as
 given in Eq.(\ref{T_c_L}). However this method has quite a large
 error. A more accurate estimate can be obtained from the
 scaling behavior of the magnetization: $m(t,L)=L^{x_m}
 \tilde{m}(tL^{1/\nu})$.  Here we set $x_m=d-d_f$ from the previous
 calculation and from an optimal scaling collapse as shown in Fig.
 \ref{Abb8} we have obtained $\nu=.73(2)$. (The area of the collapse
 region as a function of $\nu$ is shown in the inset of Fig.
 \ref{Abb8}.) Note that $\nu$ satisfies the rigorous bound for
 disordered systems\cite{ccfs}: $\nu \ge 2/d$.

\begin{figure}
  \begin{center}
     \includegraphics[width=2.35in,angle=270]{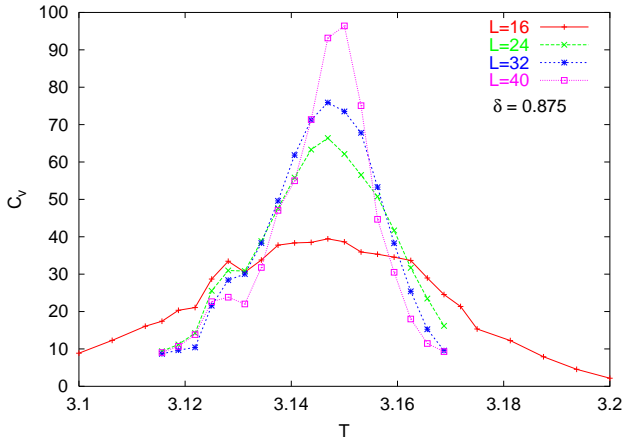}
   \end{center}
   \caption{Finite-size specific heat as a function of temperature at
     $\delta=0.875$. According to hyperscaling the maxima of the
     curves should approach a finite limiting value. The relatively large
errors are due to numerical derivation of a piece-wise linear function }
   \label{Abb9}
 \end{figure}

 Finally, we have investigated the behavior of the specific heat at
 the transition point. As seen in Fig. \ref{Abb9} the maximum of the specific
 heat is increasing with the size, therefore from the finite-size
 scaling result: $C_v^{sing}(L) \sim L^{\alpha/\nu}$ one would
 conclude $\alpha > 0$. This is, however, in conflict with
 hyperscaling and with the bound of $\nu$ in disordered systems.
 Therefore we tried to fit the numerical date by including a constant
 in the r.h.s. of the finite-size scaling form. The fit in this way,
 however, is not satisfactory. In order to get a non-positive
 $\alpha$, one should have a constant, which is more than one order of
 magnitude larger, than the finite-size data. Therefore we have
 concluded that the asymptotic regime of the specific heat is very far
 from the possibilities of present day numerical calculations.

\subsection{Tricritical transition}
\label{sec_tricritic}
\begin{figure}
  \begin{center}
     \includegraphics[width=2.35in,angle=270]{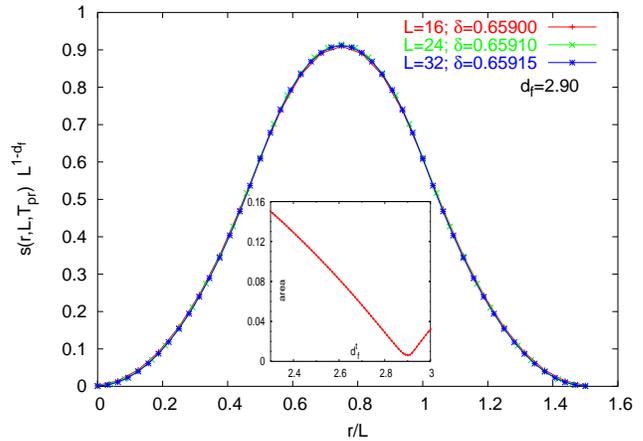}
   \end{center}
   \caption{Scaling plot of the mass of a shell of the infinite cluster analyzed along the percolation
transition temperature, Eq.(\ref{T_pr}). For each size we fix the disorder and thus the temperature
as the average spanning temperature, see text. Optimal collapse is obtained with a fractal dimension,
$d_f^t=2.94$.
Inset: area of the collapse region as a function of $d_f^t$.}
   \label{Abb10}
 \end{figure}

Analyzing the structure of the optimal set in the ordered phase we
have got a relation between the tricritical disorder and the
percolating disorder, as written in Eq.(\ref{delta_t}). Later we have
  conjectured that in this relation probably the equality holds, i.e.
  at the tricritical point three regions (disordered phase, ordered
  phase and percolating regime) meet. Here we check this conjecture
  numerically in the following way. We sit on the analytical
  continuation of the percolation transition line in Eq.(\ref{T_pr})
  for $\delta<\delta_{pr}$ (thus for $T>T_{pr}$) and calculate the
  finite-size transition point, which is located at such $\delta_t(L)$
  for which in half of the samples the giant connected cluster
  percolates the finite cube. Then we have analyzed the fractal
  properties of the giant connected cluster and repeated the procedure
  as described in Sec.\ref{sec_critic} for the second-order transition
  regime. Using the relation about the number of sites in a shell in
  Eq.(\ref{s}) and fixing the temperature for each size through
  $\delta_t(L)$ we have made a scaling analysis as in Fig.\ref{Abb7}
  for the second-order transition regime. According to the results in
  Fig.\ref{Abb10} the giant cluster is a fractal and the best collapse of data is
  obtained (see inset) with $d_f^t=2.90(2)$. This value is definitely
  different from that at a first-order transition, $d_f=d=3$, and also
  differs from that in the second-order transition regime. 
  Thus our numerical results are in accordance with the conjecture, that 
  the tricritical disorder is given by $$\delta_t = \delta_{\rm pr}$$
  Our numerical data are very sensitive to the value of $\delta$, and
  by extrapolating the finite-size results obtained from the best collapse we
find $\delta=0.65930(5)$ and $T=3.0221(1)$.
Furthermore, we obtain the estimate for the anomalous dimension of the
tricritical magnetization, $x_m^t=0.10(2)$.

\begin{figure}
  \begin{center}
     \includegraphics[width=2.35in,angle=270]{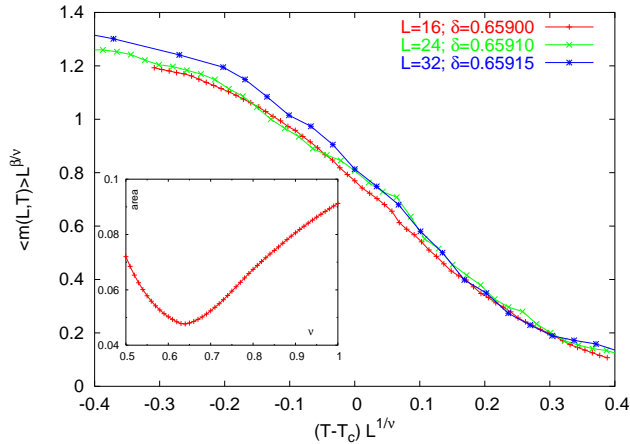}
   \end{center}
   \caption{Scaling plot of the magnetization at the finite-size tricritical
disorder as a function of the distance from the tricritical temperature, see text.
By fixing $\beta^t/\nu=d-d_f^t=2.90$ the minimum of the collapse region (see inset)
is obtained with $\nu=.64$.
  }
   \label{Abb11}
 \end{figure}

 In order to calculate the correlation length exponent, $\nu$, at the
 tricritical point we consider the magnetization, $m(t,\tau,L)$, as a
 function of reduced temperature, $t$, difference of the tricritical
 disorder, $\tau=(\delta_t-\delta)/\delta_t$, and the size, $L$.
 According to finite-size scaling we have $m(t,\tau,L)=L^{-x_m^t}
 \tilde{m}(\tau L^{1/\nu_{\tau}},t L^{1/\nu})$ and by having a
 disorder $\delta_t(L)$, (thus $\tau(L)$) for each finite size we fix
 asymptotically the first argument of the scaling function.
 Consequently form the optimal scaling collapse of
 $m(t,\tau(L),L)L^{x_m^t}$ we can obtain the correlation length
 exponent. We noticed that at the tricritical point the estimate
for $\nu$ is more sensitive to the range in which the collapse is performed.
Using the symmetric region indicated in Fig. \ref{Abb11} the minimum
of the scaling area in the inset of Fig. \ref{Abb11} is at $\nu=0.64$.
However using an asymmetric window, $-0.1 < (T-T_c)L^{1/\nu} < 0.3$, we obtain a value
which is slightly larger than the borderline value $\nu=2/3$ according
to the criterion by Chayes et al\cite{ccfs}. Therefore we conclude the estimate
of the correlation length exponent at the tricritical point as: $\nu=0.67(4)$.

 The tricritical fixed point of the random bond Potts model in the
 large-$q$ limit is related to the critical fixed point of the
 random-field Ising model according to a mapping due to Cardy and
 Jacobsen\cite{pottstm}. As described in Ref.[\onlinecite{pottstm}] the
 interface Hamiltonian of the two problems in the solid-on-solid
 approximation are equivalent to each other in $d=2+\epsilon$
 dimension, and this mapping is expected to hold for larger values of
 $d$, in particular for $d=3$.  According to the mapping
 magnetization-like excitations of the RFIM correspond to energy-like
 excitations in the random bond Potts model. In particular the
 correlation-length exponent at the tricritical point of the RBPM is
 conjectured to relate on the critical exponents of the RFIM as
\be
\nu=\nu^{RF}/(\beta^{RF}+\gamma^{RF})\;.
\label{exp_rel}
\ee
For the 3d RFIM with Gaussian disorder recent estimates on the
critical exponents are\cite{middleton}:
\be
\nu^{RF}=1.37(9),\quad \beta^{RF}=0.017(5),\quad \gamma^{RF}=2.04(14)\;,
\label{exp_RFIM}
\ee
which imply through Eq.(\ref{exp_rel}) $\nu=0.67(2)$, which coincides with our
 direct calculation.  In this way our study of the tricritical
 singularities lends support to the mapping by Cardy and Jacobsen. The
 tricritical magnetization exponent, $x_m^t$, we calculated above can
 not be predicted by the mapping therefore it gives a completely new
 piece of information.

Finally, we note that in the spirit of the RG description the
number of Potts-states, more precisely $\ln q$, is a dangerous
irrelevant scaling variable. Therefore the calculated tricritical
exponents are expected to be $q$-independent and probably universal
for any three-dimensional systems in which the first-order transition
in the pure systems is soften by tricritical disorder.
As a matter of fact, it is much easier to compute the free energy
of the Potts model at $q$ infinite, using submodular functions theory, 
than at $q$ very large. This will make  difficult to check
the assumption above, at least numerically and 
in the framework of the Potts model.

\section{Discussion}
\label{sec_disc}
Here we discuss some aspects and possible extensions of our results obtained in the previous
Sections.

\subsection{Dynamical behavior for weak disorder}
\label{griffiths}
In a random ferromagnetic system having a second-order transition
point different dynamical quantities (susceptibility, autocorrelation
function, etc.) are singular outside the critical point and this
regime is called the disordered and ordered Griffiths phases\cite{griffiths}.  The
singular behavior is due to rare regions which, due to strong disorder
fluctuations, are locally in the non-stable thermodynamic phase of the
system. It is easy to see that similar effects can be observed outside
and at a first-order transition point, as we show in the following for our
model with weak disorder. Note, however, that our numerical results
concern $q$ infinite, for which no dynamics is defined. Indeed,
only the spin-state compatible with the diagram(s) maximizing the 
function in Eq.(\ref{eq:kasfor1}) are possible, all these states being equiprobable.
The analysis below refers to $q$ very large, but finite. 

In the disordered Griffiths phase of the random bond Potts model such
a rare region is a domain (cube) of linear size $l$, having only
strong couplings, as described in Sec.\ref{connected}.  These regions
are indeed rare since they appear with a probability: $P(l) \sim
2^{3(l^3-l^2)}$. In such a cluster in equilibrium all the spins are
typically in the same state, provided the conditions in
Sec.\ref{connected} are fulfilled. However during a relaxation process
they flip into another parallel configuration. Having heat-bath
dynamics the relaxation time, $t_r$, can be estimated in such a way,
that in a thermally activated process the cluster should overcome an
energy barrier, $\Delta E(l)$, which is the energy of creation an
interface in the system. The relaxation time is then given by: $t_r
\sim \exp( \beta\Delta E(l))$.  Generally the interfacial energy is
proportional to the number of sites involved in the interface:
$\Delta(l) \sim l^{d-1} \xi_{\perp}$, where $\xi_{\perp}$ is the width
of the interface.  It is known\cite{mefisher} that outside the
first-order transition point $\xi_{\perp}$ is finite, whereas at the
transition point it is divergent as $\xi_{\perp} \sim l^{\zeta}$ with
a thermal wandering exponent: $\zeta=(3-d)/2$. In particular at $d=3$
we have a logarithmic divergence.  Now calculating the distribution of
relaxation times we obtain $P(t_r) \sim \exp[-cst~(ln
t_r)^{d/\omega})$, where $\omega=d-1$, outside the transition point of
the cluster and $\omega=d-1+\zeta$, close to the transition point.
Then the average autocorrelation function is given by:
\begin{equation}
G(t) \sim \int {\rm d} t_r P(t_r) \exp(-t/t_r) \sim \exp\left[-cst~(ln t)^{d/\omega}\right]\;,
\label{auto1}
\end{equation}
which has a different form at the first-order transition point and in the
disordered phase.

\subsection{Effect of the form of disorder}

In this paper in the numerical studies we considered bimodal i.e.
discrete form of disorder. This was mainly due to technical
simplifications.
The basic behavior of the system in particular its
critical properties are not expected to change using a continuous form
of disorder. For example the phase-diagram has the same topology as
shown in Fig.\ref{Abb2}, i.e. disordered and ordered phases and first-
and second-order transition regions. The percolation region of the
ordered phase also exists and very probably the tricritical point is
located at the meeting point of these three phases or regions. Also
the critical singularities at the second-order transition line are
very probably disorder independent, as we have already observed in the
two-dimensional problem\cite{long2d}. To the same question for the
tricritical transition is more difficult to answer. Having in mind
that the $3d$ RFIM might have disorder dependent
singularities\cite{sourlas} the same can be true for the tricritical
singularities, due to the mapping as described in
Sec.\ref{sec_tricritic}. Finally the discrete nature of the disorder
can result in non-physical discontinuities in the internal energy,
which are washed out by continuous disorder.

\subsection{Effect of the number of states - $q$}

In the second-order transition regime the critical exponents are
$q$-dependent, which can be seen from the results of numerical
investigations\cite{pottssite,pottsbond} and the same scenario holds
in two dimensions, too. The tricritical singularities, however, at
least those which are related to the energy density, are very probably
$q$-independent and thus 'hyperuniversal'. It would be very interesting
to check this statement for another systems,
since the Potts model in three dimensions for $q$ finite
but very large, seems out of the reach of the usual methods.
 Finally we mention that
the non-physical discontinuities in the internal energy are also
absent for finite value of $q$.

\subsection{Model with correlated disorder}

Finally, we consider our model with correlated disorder, in which
translational invariance is present in the vertical direction, in
which the couplings are constant, $J_{\perp}$, whereas in the
horizontal $2d$ planes the couplings are random, $J_{ij}$, and
strictly correlated in each planes. This type of columnar disorder is
introduced by McCoy and Wu in the $2d$ Ising model\cite{mccoywu}. Due to
translational symmetry in the vertical direction the model is
conveniently studied in the transfer matrix formalism. Using the
extreme anisotropic limit of the model\cite{kogut}, when, $J_{\perp}/J_{ij} \to 0$
the transfer matrix is written into the form: $\cal{T}=\exp(-\tau
\cal{H})$, where $\tau$ is the infinitesimal lattice spacing and
$\cal{H}$ is the Hamiltonian operator of the $2d$ quantum Potts model
with random couplings. This latter model can be studied by a strong
disorder renormalization group method\cite{review} which leads to $q$-independent
critical properties. According to numerical results\cite{2drg,review} the correlation
length critical exponent is $\nu=1.15(10)$ and the anomalous dimension
of the magnetization is $x_m^{an}=0.97(3)$. This latter result implies
that the fractal dimension of the giant anisotropic cluster is given
by: $d_f^{c}=d-x_m^{c}=2.03(3)$, and evidently $d_f^{c} < d_f$.
Starting with the anisotropic model we can go to the isotropic model
by letting the couplings to be random in the vertical direction, too.
Our results indicates that during this process the mass of the largest
cluster is increasing, i.e. the creation of new connected parts is
more effective than the creation of isolated sites. This result is in
accordance with the form of the phase-diagram in Fig. \ref{Abb2}, in
which due to a similar process the ordered phase has a larger extent
with increasing disorder. We note that in $2d$ our numerical results
show\cite{long2d} that $d_f^{an} = d_f$, which let us to conjecture the exact
values of the critical exponents in the random bond Potts model. Here
we have argued that in $2d$ due to duality the creation and
destruction processes play equivalent r\^ole and therefore the fractal
dimension stays unchanged.

This work has been supported by the French-Hungarian cooperation
programme Balaton (Minist\'ere des Affaires Etrang\`eres - OM), the
Hungarian National Research Fund under grant No OTKA TO34183, TO37323, TO48721,
MO45596 and M36803. F.I. thanks for useful discussions with C. Monthus and H. Rieger.

\end{document}